\documentclass[dvips]{acta74}
\usepackage{supertabular,lscape,epsfig}
\usepackage{amssymb}
\usepackage{amsmath}

\SetPages{1}{1}

\SetVol{74}{2024}

\begin{document}

\begin{Titlepage}
\Title{Red giant component of the recurrent nova T Coronae Borealis}
\Author{ R. K. Zamanov$^1$, D. Marchev$^2$, V. Marchev$^1$, M. Moyseev$^1$, \\
         G. Yordanova$^2$, J. Mart\'i$^3$, M. F. Bode$^{4,5}$, \& K. A. Stoyanov $^1$}
{$^1$Institute of Astronomy and National Astronomical Observatory, Bulgarian Academy of Sciences,
     Tsarigradsko Shose 72, BG-1784 Sofia, Bulgaria  \\
 $^2$Department of Physics and Astronomy, Shumen University "Episkop Konstantin Preslavski",         
     115 Universitetska Str., 9700 Shumen, Bulgaria  \\
 $^3$Departamento de F\'isica, Escuela Polit\'ecnica Superior de Ja\'en, Universidad de Ja\'en, 
     Campus Las Lagunillas s/n,  A3-420, 23071, Ja\'en, Spain \\
 $^4$Astrophysics Research Institute, Liverpool John Moores University, IC2, 149 Brownlow Hill, 
    Liverpool, L3 5RF, UK \\ 
 $^5$Office of the Vice Chancellor, Botswana International University of Science and Technology, 
    Private Bag 16, Palapye, Botswana          
}

\Received{October 3, 2024}
\end{Titlepage}

\Abstract{We performed simultaneous V band photometry 
and spectroscopic observations of the recurrent nova 
T~CrB and estimate the V band magnitude of the red giant.  
We find for the red giant of T~CrB  
apparent and absolute V-band magnitudes $m_V = 10.17 \pm 0.06 $ and 
$M_V = +0.14 \pm 0.08$, respectively.
At the maximum of the ellipsoidal variation when these values are obtained, 
its absolute V-band magnitude is similar but fainter than the typical M4/5III giants. \\
The data are available on :  zenodo.org/records/15174720 
}
{Stars: binaries: symbiotic -- accretion, accretion discs -- novae, cataclysmic variables 
         -- stars: individual: T~CrB }

\section{Introduction}

T Coronae Borealis (HD 143454, NOVA CrB 1946, NOVA CrB 1866) is a famous recurrent nova
having recorded eruptions in 1946, 1866, 1787 (Schaefer 2023a), 
when the star peaked at $\sim 1.7$ magnitude as  observed by 
A.S. Kamenchuk, M. Woodman, N.F.H. Knight (see Kukarkin 1946; Shears 2024a). 
The recurrence time of the outbursts is of about 80 yr, and 
a new eruption  is thought likely to occur in the next months (Luna et al. 2020; Schaefer 2024b;
Shears 2024b).

At quiescence T~CrB shows  (1) an optical spectrum with M-type absorption 
features with the additional HI, HeI, HeII, and [OIII] emission lines 
(Kenyon 1986; Iijima 1990; Munari et al. 2016) 
and (2) behaviour of a dwarf nova with an extremely long orbital period (I{\l}kiewicz et al. 2023). 
The binary  consists  of a M4III red giant  
(Kenyon \& Fernandez-Castro 1987; M{\"u}rset \& Schmid 1999) 
and a massive 1.2-1.4~M$_\odot$ white dwarf   
(Belczynski \& Mikolajewska 1998; Stanishev et al. 2004). 
The orbital period of the system is 227.5687~d  
(Fekel et al. 2000). 


Here we  report simultaneous V-band photometry and spectroscopic observations, 
performed with aim of estimating the V-band magnitude of the red giant.  

\section{Observations}

During the night of 22/23 August 2024, 
T~CrB was observed  simultaneously with the 2.0~m Rozhen  
and with the 40~cm Shumen telescopes. 
Optical spectra of T~CrB and of four red giants were secured with the 
ESpeRo Echelle spectrograph (Bonev et al. 2017) 
on the 2.0~m~RCC telescope in the Rozhen  National Astronomical Observatory, Bulgaria.
The spectrograph covers the range between 3900 -- 9000~\AA\ with a resolution reaching 45~000 around the H$\alpha$ line. The usefull spectral range for T~CrB is from 4600~\AA\ to 6800~\AA.
Simultaneously with the spectral observations, 
T~CrB was monitored in UBV bands with the 40~cm telescope of the Shumen
University "Episkop Konstantin Preslavski" (Kjurkchieva et al. 2020). 
The spectra and the photometry were processed with IRAF. 
For the photometry, comparison stars from the list of Henden \& Munari (2006) were used.

The photometric observations are presented in Table~1.
In this table are given UT, number of the data points, minimum, maximum and average magnitude, standard 
deviation and the typical observational error.   
The spectroscopic observations are summarized in Table~2. 
This table lists object, its spectral type, UT of the start of the exposure, exposure time in minutes, 
signal-to-noise ratio at 5500~\AA. 

\MakeTable{lclc| r  r |r| c r r }{12.5cm}{Photometric observations of T~CrB. The date is in the format yyyy-mm-dd.}
{\hline
date       & UT start-end  &      &  N$_{pts}$ & min    & max    & average  & stdev &  merr  \\
           & hh:mm - hh:mm &      &            & [mag]  & [mag]  & [mag]    & [mag] & [mag]  \\
\hline
           &               &      &            &        &        &          &       &        \\
  2024-08-22 & 20:03 - 21:28 & U &  23x120s & 11.452 & 11.899 & 11.558   & 0.109 &  0.060 \\
  2024-08-22 & 20:05 - 21:29 & B &  22x40s & 11.146 & 11.229 & 11.186   & 0.021 &  0.008 \\
  2024-08-22 & 20:05 - 21:30 & V &  22x15s &  9.916 &  9.974 &  9.950   & 0.019 &  0.004 \\
\hline
}

\MakeTable{c c c c | c |c c }{12.5cm}{Spectral  observations of T~CrB and comparison red giants.}
{\hline
object    & Spec.      & date          & UT start & exp-time & S/N & \\
          &            & yyyy-mm-dd    & hh:mm    & [min]    &     & \\
\hline
          &            &            &       &     &       \\
T~CrB     & M4.5III+WD & 2024-08-22 & 20:12 &  10 &  17 &  spec.1\\
          &            & 2024-08-22 & 20:23 &  60 &  35 &  spec.2\\
          &            &            &       &     &       \\
NV~Peg    & M4.5III    & 2024-08-22 & 21:31 &  15 & 128 & \\
          & 	       & 2024-08-22 & 21:47 &  15 & 127 & \\
          &            &            &       &     &       \\
XZ~Psc    &  M5III     & 2024-08-23 & 00:10 &  15 & 107 & \\
          &            & 2024-08-23 & 00:26 &  15 & 111 & \\
          &            & 2024-08-23 & 00:42 &   5 &  64 & \\
          &            &            &       &     &       \\
71~Peg    &  M5III     & 2024-08-23 & 00:52 &   5 & 68  & \\
          &	       & 2024-08-23 & 00:58 &  15 & 117 & \\
          &            & 2024-08-23 & 01:14 &  15 & 118 & \\
          &            &            &       &     &       \\
57~Psc    &  M4III     & 2024-08-23 & 02:10 &  15 & 120 & \\
          &            & 2024-08-23 & 02:04 &   5 & 71  & \\
          &            &            &       &     &       \\
\hline }

\MakeTable{rc | c  c | c c r r }{7.5cm}{Contribution of the red giant of T~CrB. }
{\hline
                 & red giant & 5420 - 5460 \AA  & 5480-5530 \AA      &  \\
\hline

T~CrB  spec.1    & NV~Peg    & $0.85 \pm 0.02$  &    ---             & \\
T~CrB  spec.2    & NV~Peg    & $0.89 \pm 0.01$  &    ---             & \\ 
                 &           &                  &                    & \\
T~CrB  spec.1    & XZ~Psc    & $0.76 \pm 0.02$  &   $0.77 \pm 0.03$  & \\
T~CrB  spec.2    & XZ~Psc    & $0.80 \pm 0.01$  &   $0.78 \pm 0.02$  & \\
                 &           &                  & 		     & \\
T~CrB  spec.1    & 71~Peg    & $0.78 \pm 0.02$  &   $0.76 \pm 0.03$  & \\
T~CrB  spec.2    & 71~Peg    & $0.81 \pm 0.01$  &   $0.83 \pm 0.02$  & \\
                 &           &                  & 		     & \\         
T~CrB  spec.1    & 57~Psc    & $0.87 \pm 0.02$  &   $0.79 \pm 0.03$  & \\
T~CrB  spec.2    & 57~Psc    & $0.90 \pm 0.01$  &   $0.86 \pm 0.02$  & \\
                 &           &                  &  		     & \\
\hline }

 \begin{figure*}     
   \vspace{5.9cm}   
   \includegraphics{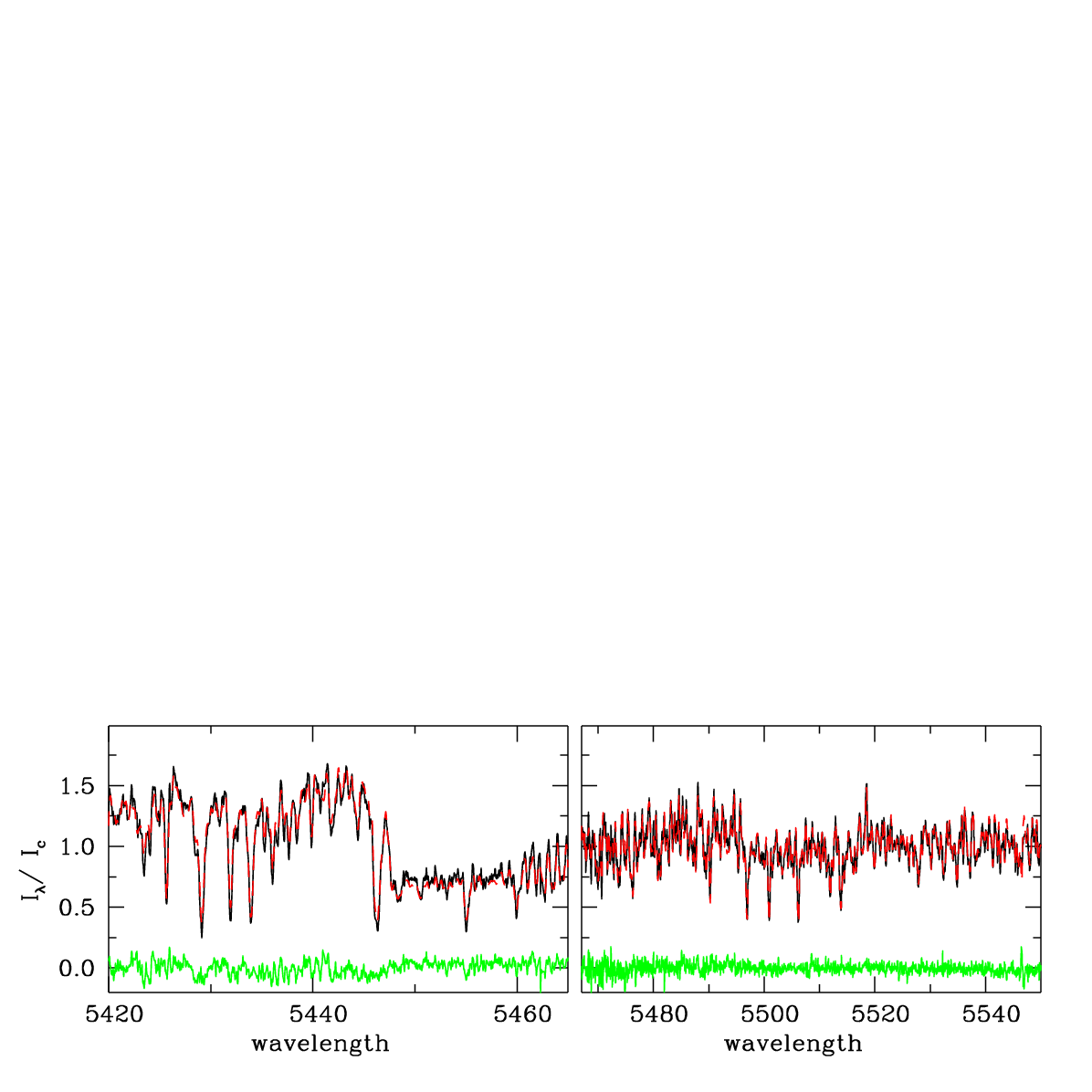}	 
   \caption[]{T CrB (black solid line),  the fit (red dashed line), 
   and the residuals (green). 
   On the left panel, 71~Peg is used (83\% contribution of the giant). 
   On the right panel, 57~Psc is used (88\% contribution of the giant). }
  \label{f1}      
 \end{figure*}	     


M{\"u}rset \& Schmid (1999) classified the cool component of T~CrB
as M4.5III. From the Yale Bright Star Catalog (Hoffleit  \& Warren 1995)
we selected 4 red giants of similar spectral type and observed them 
with the same setup in the same night. 
Their  spectral classifications from 
the 14th General Catalogue of MK Spectral Classification
(Buscombe \& Foster 1999)
and the Extended Hipparcos Compilation (Anderson \& Francis 2012) 
are similar:  
{\bf (1) NV~Peg} is classified as M4.5III-IIIa (Hoffleit  \& Warren 1995),
M4.5IIIA (Buscombe \& Foster 1999), and M4.5IIIa (Anderson \& Francis 2012);
{\bf (2) XZ~Psc} -- M5III (Hoffleit  \& Warren 1995), M4.6III (Buscombe \& Foster 1999),
             M3/4~III (Anderson \& Francis 2012);
{\bf (3) 71~Peg} -- M5IIIa (Hoffleit  \& Warren 1995), M4.7IIIA (Buscombe \& Foster 1999);
and M4III (Anderson \& Francis 2012);
{\bf (4) 57~Psc} -- M4IIIa (Hoffleit  \& Warren 1995),  M4IIIA (Buscombe \& Foster 1999),
and M4III (Anderson \& Francis 2012). 

%
%
%

\section{Contribution of the red giant}

The photometric V-band has effective wavelength 5445.43~\AA\ 
and central wavelength 5512.10~\AA\  (e.g. Rodrigo \& Solano 2020). 
We selected two regions on our spectra around these wavelengths 
to calculate the contribution of the red giant. 
We used the following procedure: 
(1) we normalize  the spectra to the average value (the local continuum); 
(2) we multiply the spectrum of the red giant 
with a factor from 0.05 to 1.00 with a step of 0.01; 
(3) we subtract the red giant contribution from the spectra of T~CrB;
(4) we find the value of the scaling factor that produces the minimum of the standard deviation 
of the residuals.  
This minimum represents the (fractional) contribution of the red giant.
This is done for 5420-5460~\AA\  and range 5480-5530~\AA, for the two spectra of T~CrB, 
as well as for each of the four red giants listed in Table~2.
The results are summarized  in Table 3.


%
%

We estimated for the first spectrum 
contribution  of the red giant $0.82 \pm 0.05$ to the 
wavelength range  5420-5460~\AA\ and $0.77 \pm 0.02$ to the  5480-5550~\AA. 
For the second  spectrum the contribution of the red giant is
$0.85 \pm 0.05$ and $0.84 \pm 0.05$, respectively. 
Using the simultaneous V-band photometry: 
(1) for the first spectrum we estimate an average V-band band magnitude 
of T~CrB  $9.953 \pm 0.015$, contribution of the red giant $80\% \pm 4 \%$, and 
apparent magnitude  of the red giant  $m_V = 10.20 \pm 0.07$;
(2) for the second spectrum -- average V-band magnitude 
of T~CrB  $9.950 \pm 0.014$, contribution of the red giant $84\% \pm 5 \%$, and 
apparent magnitude  of the red giant  $m_V = 10.14 \pm 0.06$.
The overall result from the two spectra is $m_V = 10.17 \pm 0.06$.

We adopt a distance to T~CrB of $d=914$~pc (Schaefer 2022), 
which is similar to the value $d =890$~pc (Bailer-Jones et al. 2021) based on the Gaia EDR3
(Gaia Collaboration 2021).  We also adopt interstellar extinction
$E_{B-V}$ = 0.07 (Nikolov 2022). This value is consistent with
the Galactic dust reddening maps by Schlegel et al. (1998) and
Schlafly \& Finkbeiner (2011), which give an upper limit $E_{B-V} \le 0.071$ 
(calculated with the NASA/NED extinction calculator). \\
Using  $M_V = m_V - 3.1 E_{B-V} - 2.5 \log [(d/10)^2]$, we estimate the
absolute V-band  magnitude of the red giant of T~CrB as
$M_V=+0.14 \pm 0.08$. 
This value refers to our 22 August 2024 observations, 
which are on orbital phase  0.49  
(calculated with the ephemeris $P_{orb}= 227.5687$~d, $JD_0 = 2447918.62$ 
by Fekel et al 2000).

\section{Discussion} 

In the recurrent nova T~CrB the  red giant is  ellipsoidally shaped.
The ellipsoidal variability was first demonstrated by Bailey (1975).
Later, Lines et al. (1988) analysed the ellipticity effect at UBVRI bands 
and Yudin  \&  Munari  (1993) -- in J band. 
The red giant is stable - its V-band light curve 
has not changed in its main features over two decades (Zamanov et al. 2004). 
The stability of the red giant is better defined 
in the IR observations  (Yudin \& Munari 1993, Shahbaz et al.,1997), 
where an upper limit of variability $\Delta J <0.02$ has been constrained.
The red giant fills its Roche lobe and transfers material
via the Lagrangian point L$_1$ at a rate $\sim 10^{-8}$~M$_\odot$ yr$^{-1}$
(Selvelli et al. 1992;  Zamanov et al. 2023).

We find $M_V=+0.14 \pm 0.08$ on orbital phase 0.49, 
close to the maximum of the ellipsoidal variation. 
The  amplitude of the ellipsoidal variation is 0.4 mag, and 
at the minimum it would  be  $M_V \sim +0.55$~mag.
Based on previous estimations the value of the absolute magnitude of M4III giant would be an $M_V = -0.6$ and for an M5III star $M_V = -0.1$(e.g. Table 2 in Straizys \& Kuriliene 1981). . 
It is worth noting that other studies give slightly brighter values:  
 $M_V = -0.94$ for M4III and  $M_V = -0.69$ for  M5III  (Th\'e et al. 1990); 
 $M_V = -0.5$  for M4III and  $M_V = -0.3$  for  M5III  (Schmidt-Kaler  1982);
 $M_V = -1.0$  for M4III and  $M_V = -0.7$  for  M5III  (Mikami \& Heck 1982). 
 
Our result has a lower value than in the literature so far for such giants. 
Kenyon \& Fernandez-Castro (1987) on the basis of TiO, VO and NaI infrared doublet
classified the cool component of T~CrB as  M$4.1 \pm 0.3$III. 
M{\"u}rset \& Schmid (1999) using TiO band head $\lambda 8432$ (see their Table~3) find
M4.5III -- M5III.  Our results indicate that even at the maximum of the ellipsoidal variations, 
its absolute V-band magnitude is fainter than that of an average M4III-M5III giant. 
This is probably a result of the evolution of the binary system (e.g. Chen et al. 2010), that
produced a red giant with a smaller radius and/or lower mass compared to the average value of a
giant of the same spectral class -- a result of the mass loss towards the compact object and that the red giant is confined (restricted) by the Roche lobe.  


The estimated magnitudes of the red giant could be useful:  
(1) for disentangling the composite spectrum of the system (e.g. Skopal  2005)
    before, during and after  the nova outburst; 
 (2) to model the ellipsoidal variability (e.g. Belczynski \& Mikolajewska 1998); 
(3) to study possible changes in the red giant --
its  atmosphere is expected to be ionized by the nova explosion (e.g. Page et al. 2020);
and (4) the irradiation by the cooling white dwarf during the secondary maximum  (Munari 2023). 

\section{Conclusions}

We performed simultaneous spectral (2.0~m Rozhen telescope) 
and photometric (40~cm Shumen telescope) observations
of the recurrent nova T~CrB on 22  August 2024. 
Using spectra of red giants 
of M4III-M5III spectral types obtained in the same night,  
we estimate for the red giant of T~CrB 
apparent and absolute V-band magnitudes $m_V = 10.17 \pm 0.06 $ and 
$M_V = +0.14 \pm 0.08$, respectively.

\Acknow{
We acknowledge project PID2022-136828NB-C42 funded
by the Spanish MCIN/AEI/ 10.13039/501100011033, "ERDF A way of making Europe",
Bulgarian Ministry of Education and Science (the National Program Young Scientists and Postdoctoral Students-2), as well as by project RD-08-137/2024 of Shumen University. 
We thank an anonymous referee for some very insightful 
and helpful comments on the initial manuscript.}

\end{document}